\documentstyle[psfig]{mn}

\newif\ifAMStwofonts

\ifoldfss
  \ifCUPmtlplainloaded \else
    \NewTextAlphabet{textbfit} {cmbxti10} {}
    \NewTextAlphabet{textbfss} {cmssbx10} {}
    \NewMathAlphabet{mathbfit} {cmbxti10} {} 
    \NewMathAlphabet{mathbfss} {cmssbx10} {} 
  \fi
  \ifAMStwofonts
    \ifCUPmtlplainloaded \else
      \NewSymbolFont{upmath} {eurm10}
      \NewSymbolFont{AMSa} {msam10}
      \NewMathSymbol{\upi}     {0}{upmath}{19}
      \NewMathSymbol{\umu}     {0}{upmath}{16}
      \NewMathSymbol{\upartial}{0}{upmath}{40}
      \NewMathSymbol{\leqslant}{3}{AMSa}{36}
      \NewMathSymbol{\geqslant}{3}{AMSa}{3E}

       \let\le=\leqslant
       
    \fi
  \fi
\fi 

\ifnfssone
  \newmathalphabet{\mathit}
  \addtoversion{normal}{\mathit}{cmr}{m}{it}
  \addtoversion{bold}{\mathit}{cmr}{bx}{it}
  \newmathalphabet{\mathbfit} 
  \addtoversion{normal}{\mathbfit}{cmr}{bx}{it}
  \addtoversion{bold}{\mathbfit}{cmr}{bx}{it}
  \newmathalphabet{\mathbfss} 
  \addtoversion{normal}{\mathbfss}{cmss}{bx}{n}
  \addtoversion{bold}{\mathbfss}{cmss}{bx}{n}
  \ifAMStwofonts
    \ifCUPmtlplainloaded \else
      \UseAMStwoboldmath
      \makeatletter
      \new@mathgroup\upmath@group
      \define@mathgroup\mv@normal\upmath@group{eur}{m}{n}
      \define@mathgroup\mv@bold\upmath@group{eur}{b}{n}
      \edef\UPM{\hexnumber\upmath@group}
      \new@mathgroup\amsa@group
      \define@mathgroup\mv@normal\amsa@group{msa}{m}{n}
      \define@mathgroup\mv@bold\amsa@group{msa}{m}{n}
      \edef\AMSa{\hexnumber\amsa@group}
      \makeatother
      \mathchardef\upi="0\UPM19
      \mathchardef\umu="0\UPM16
      \mathchardef\upartial="0\UPM40
      \mathchardef\leqslant="3\AMSa36
      \mathchardef\geqslant="3\AMSa3E

       \let\le=\leqslant

    \fi
  \fi
\fi 

\ifnfsstwo
  \DeclareMathAlphabet{\mathbfit}{OT1}{cmr}{bx}{it}
  \SetMathAlphabet\mathbfit{bold}{OT1}{cmr}{bx}{it}
  \DeclareMathAlphabet{\mathbfss}{OT1}{cmss}{bx}{n}
  \SetMathAlphabet\mathbfss{bold}{OT1}{cmss}{bx}{n}
  \ifAMStwofonts
    \ifCUPmtlplainloaded \else
      \DeclareSymbolFont{UPM}{U}{eur}{m}{n}
      \SetSymbolFont{UPM}{bold}{U}{eur}{b}{n}
      \DeclareSymbolFont{AMSa}{U}{msa}{m}{n}
      \DeclareMathSymbol{\upi}{0}{UPM}{"19}
      \DeclareMathSymbol{\umu}{0}{UPM}{"16}
      \DeclareMathSymbol{\upartial}{0}{UPM}{"40}
      \DeclareMathSymbol{\leqslant}{3}{AMSa}{"36}
      \DeclareMathSymbol{\geqslant}{3}{AMSa}{"3E}

       \let\le=\leqslant

    \fi
  \fi
\fi 

\ifCUPmtlplainloaded \else
  \ifAMStwofonts \else 
    \def\upi{\pi}
    \def\umu{\mu}
    \def\upartial{\partial}
  \fi
\fi

\title[Formation of $\omega$ Centauri]
{Formation of $\omega$ Centauri  from an ancient nucleated dwarf galaxy  
in the young Galactic disc}
\author[K. Bekki, K. C. Freeman]
       {K. Bekki,${}^1$  and K. C. Freeman${}^2$\\
        ${}^1$School of Physics, University of New South Wales, Sydney 2052, NSW, Australia \\
        ${}^2$Research School of Astronomy \& Astrophysics,
        Mt Stromlo Observatory, The Australian National University,
        Cotter Road, Weston Creek,   \\ ACT 2611, Australia}
\date{Accepted 
      Received
      in original form 2001}

\pagerange{\pageref{firstpage}--\pageref{lastpage}}
\pubyear{1994}

\begin{document}

\maketitle

\label{firstpage}

\begin{abstract}

We first present a self-consistent dynamical model in which
$\omega$ Cen is formed from an ancient 
nucleated dwarf galaxy merging with the first generation of
the Galactic thin disc in a retrograde manner with respect to the Galactic rotation.  
Our numerical simulations 
demonstrate  that  during merging between the Galaxy and the $\omega$ Cen's host 
dwarf  with $M_{\rm B}$ $\simeq$  $-14$ mag and its nucleus mass of $10^7$ $M_{\odot}$,
the outer stellar envelope of the dwarf is nearly completely stripped
whereas the central nucleus can survive from the tidal stripping
because of its compactness.
The developed naked nucleus 
has a very bound retrograde orbit around the young Galactic disc,
as observed for $\omega$ Cen,
with its apocenter and pericenter distances
of $\sim$ 8 kpc and $\sim$ 1 kpc, respectively.
The Galactic tidal force can induce  
radial inflow of gas to the dwarf's center
and consequently triggers 
moderately strong nuclear starbursts in a repetitive manner. 
This  result implies that 
efficient  nuclear chemical enrichment 
resulting from the later starbursts 
can be  closely associated with the origin 
of the observed relatively young and metal-rich stars
in $\omega$ Cen.
Dynamical heating  by the $\omega$ Cen's host can transform the young thin disc
into  the thick disc during merging. 

\end{abstract}

\begin{keywords}
globular clusters:individual ($\omega$ Centauri) 
\end{keywords}

\section{Introduction}

The most massive Galactic globular cluster $\omega$ Cen is observed
to have unique physical properties, such as a very flattened shape
for a globular cluster (e.g., Meylan 1987),  broad metallicity distribution 
(e.g., Freeman \& Rodgers 1975; Norris et al. 1996), 
strong variations of nearly all element abundances among its stars
(e.g., Norris \& Da Costa 1995; Smith et al. 2000),
kinematical difference between its metal-rich and metal-poor 
stellar populations (e.g., Norris et al. 1997),
multiple stellar populations with different spatial distributions
(e.g., Pancino et al. 2000; Ferraro et al. 2002),
star formation history extending over a few Gyr (Lee et al. 1999; Smith et al. 2000),
and its very bound retrograde orbit with respect to the Galactic rotation
(Dinescu et al. 1999).
These unique characteristics have been  considered to suggest that 
there are  remarkable differences in star formation  histories, chemical enrichment processes,
and structure formation between $\omega$ Cen  and other Galactic normal
globular clusters (e.g., Hilker \& Richtler  2000, 2002)

The observed  extraordinary nature of $\omega$ Cen has attracted much attention from
theoretical and numerical works on chemical and dynamical evolution of  $\omega$ Cen 
(e.g., Icke \& Alca\'ino 1988; Carraro \& Lia 2000; Gnedin et al. 2002; Zhao 2002).
One of the most extensively discussed scenario for $\omega$ Cen formation
is that $\omega$ Cen is the surviving nucleus of an ancient nucleated dwarf galaxy
with its outer  stellar envelope almost entirely removed by  tidal stripping
of the Galaxy 
(Zinnecker et al. 1988; Freeman 1993).
The observed atypical bimodal or multi-model metallicity distribution
(e.g., Norris et al. 1996)
and the metal-rich stellar population 2 $-$ 4 Gyr younger than the metal-poor
(Lee et al. 1999; Hilker \& Richtler 2000;  Hughes \& Wallerstein 2000)
have been suggested to support this scenario.
However, because of the lack of extensive numerical studies on
dynamical evolution of {\it nucleated}  dwarf galaxies interacting/merging  with the Galaxy,
it remains unclear when and how an ancient nucleated dwarf
galaxy lose {\it only} its stellar envelope without totally destroying its nucleus
in its dynamical interaction with the Galaxy.

The purpose of this paper is to demonstrate that 
$\omega$ Cen can be formed from an ancient nucleated dwarf galaxy
interacting/merging  with the young Galactic disc 
($\sim$ 10 Gyr ago).
We consider that if $\omega$ Cen is formed from merging between a massive,
compact  nucleated dwarf
and the Galaxy, the merging epoch should be well before the formation of
the present-day thin disc, because such a massive dwarf can significantly
heat up the thin Galactic disc (e.g., Quinn et al.  1993).
Our fully self-consistent 
numerical simulations  demonstrate  that the stellar envelope of the nucleated dwarf
with $M_{\rm B}$ $\sim$ $-14$
can be nearly completely stripped by the  strong tidal field 
of the first generation of the Galactic thin disc with the stellar mass
only $\sim$ 10 \%  of the mass of the present-day Galactic thin disc
(i.e.,  the same as that of the present-day thick disc) 
whereas the central nucleus can remain intact owing to its compactness.
Recently Mizutani et al. (2003) and Tsuchiya et al. (2003)
have discussed the formation of $\omega$ Cen in terms of tidal disruption
of a dwarf by {\it the present-day}  Galaxy.
We discuss the origin of the relatively metal-rich and young stellar populations 
of $\omega$ Cen in terms of the dwarf's star formation history
strongly influenced by tidal interaction with the first generation
of the Galactic thin disc.

\begin{figure}
\psfig{file=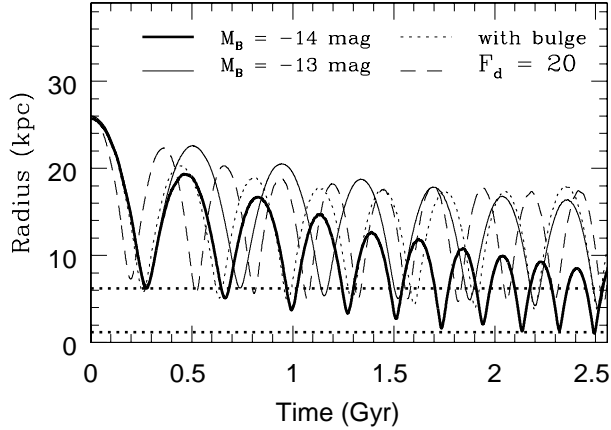,width=8.cm}
\caption{ 
Orbital evolution of the nucleated dwarf for four different models
with initial $e_{\rm p}$ $\sim$ 0.63:
The fiducial model with $M_{\rm B}$ = $-14.0$ mag, 
$F_{\rm d}$ =5, and $M_{\rm b}$ = 0
(thick solid),
less luminous one with  $M_{\rm B}$ = $-13.0$ mag,
$F_{\rm d}$ =5, and $M_{\rm b}$ = 0
(thin solid),
bulge model with  $M_{\rm B}$ = $-14.0$ mag,
$F_{\rm d}$ =5, and $M_{\rm b}$ = 10$^{10}$ $\rm M_{\odot}$ (thin dotted),
and the more massive Galaxy model with $M_{\rm B}$ = $-14.0$ mag,
$F_{\rm d}$ =20, and $M_{\rm b}$ = 0 
(dashed).
The upper and lower thick (horizontal) lines represent the observed
apocenter (6.2 kpc) and pericenter distance (1.2 kpc)  
of $\omega$ Cen's orbit (Dinescu et al. 1999).
Note that only the nucleus  of the more luminous dwarf can reach the central
region of the Galaxy within a few Gyr.
Note  also that the dwarfs in the latter two models 
can not  approach the inner region  of the Galaxy because the dwarfs 
are  completely destroyed  before dynamical friction cause significant
orbital decay of the dwarf. 
}
\label{Figure. 1}
\end{figure}

\begin{figure}
\psfig{file=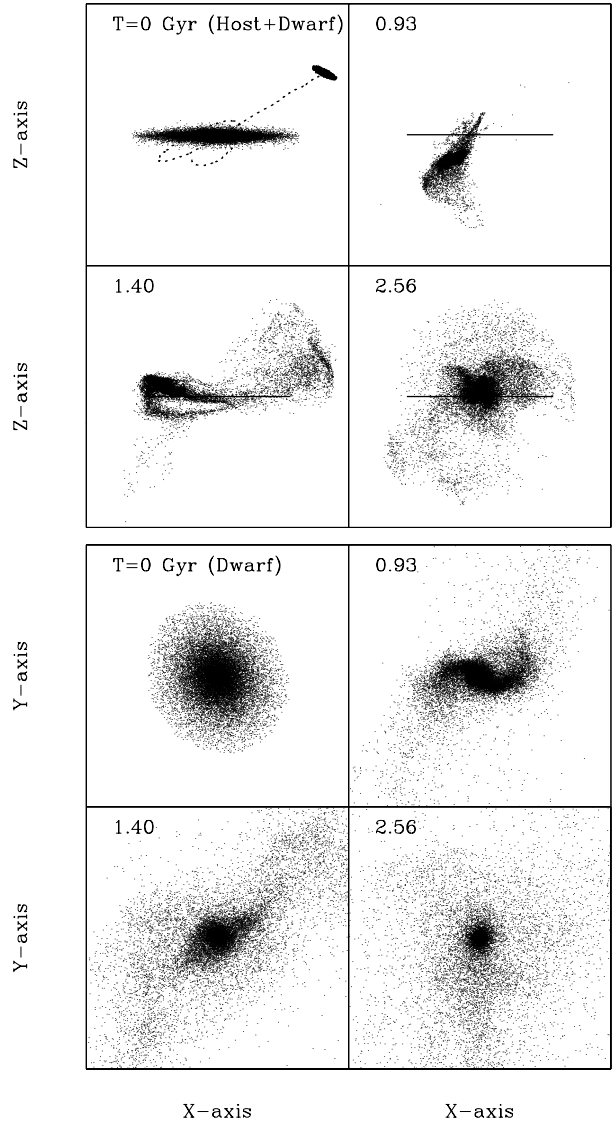,width=8.cm}
\caption{ 
Morphological evolution of stellar components of  the nucleated dwarf galaxy 
projected onto the $x$-$z$ plane (upper four) and
onto the  $x$-$y$ one (lower four) in the fiducial model.
For clarity, the Galactic plane is represented  as a solid line 
in the three of the upper four panels.  
The time $T$ (in our units) represents the time that has elapsed since
the simulation starts.
Each frame in the upper (lower) four panels
measures 54.6 (9.4) kpc on a side.
For comparison, we plot the initial Galactic disc stars at $T$ = 0 
and the orbit for the first 0.93 Gyr (dotted)
in the upper left panel of the upper four.
}
\label{Figure. 2}
\end{figure}

\section{The model}

\subsection{The Galaxy}

We construct  the dynamical model of the Galaxy embedded in a massive dark matter halo
by using Fall-Efstathiou model (1980) with the total mass of $M_{\rm t}$.
The  exponential disc of the Galaxy is 
assumed to have the radial scale length ($a_{\rm d}$) of 3.5 kpc,
the size ($R_{\rm d}$) of 17.5 kpc,
the mass of  $M_{\rm d}$, and the  ratio 
of the total Galactic mass 
to the disc mass ($M_{\rm t}/M_{\rm d}$, hereafter referred to as $F_{d}$). 
These $M_{\rm d}$ and $F_{d}$ are considered to be important
parameters for the young Galactic disc.
In the adopted  Fall-Efstathiou model,
the rotation curve becomes nearly flat at 1.75 $a_{d}$ and the dark matter halo
is truncated at 8.0  $a_{d}$, which is beyond the region
reached by adopted orbit of $\omega$ Cen.  If we assume that 
$M_{\rm d}$ = 6.0 $\times$ $10^{10}$ $ \rm M_{\odot}$
(hereafter referred to as $M_{\rm d} (0)$) 
and $M_{\rm t}$ = 3.0 $\times$ $10^{11}$ $ \rm M_{\odot}$
(hereafter $M_{\rm t} (0)$),  
the maximum circular velocity is 220 km s$^{-1}$.
In addition to the rotational velocity associated with  the gravitational
field of disc and halo component,
the initial radial and azimuthal velocity
dispersion are given to disc component according
to the epicyclic theory with Toomre's parameter (Binney \& Tremaine 1987) $Q$ = 1.5. 
The vertical velocity dispersion at given radius
are set to be 0.5 times as large as
the radial velocity dispersion at that point,
as is consistent  with
the observed trend  of the Milky Way (e.g., Wielen 1977).

In the present study, we consider that merging between 
the $\omega$ Cen's progenitor and the young Galaxy happened when the {\it stellar disc mass}
of the Galaxy is only 10 \% of the mass of the present-day Galactic thin disc
($M_{\rm d}(0)$). The {\it total mass}
of the young Galactic disc is highly likely to be 
larger than 0.1 (because of the possibly gas-rich
nature of the disc) and we mainly present the results of the
models with  $M_{\rm d}$ = 0.2 $\times$ $M_{\rm d}(0)$. 
Since there are no observational  constraints on the dark matter content
of the young galaxy, 
we mainly investigate the models with $F_{d}$ = 10.0 
($M_{\rm t}$=0.4 $\times$  $M_{\rm t}(0)$)
and 20.0 ($M_{\rm t}$ = 0.8 $\times$ $M_{\rm t}(0)$). For comparison, 
we also investigate the model  with a  Galactic bulge with the $R^{\rm 1/4}$ density
profile, the effective radius of 0.7 kpc, and the mass ($M_{\rm b}$)
of $10^{10}$ $\rm M_{\odot}$. 
 
\subsection{The $\omega$ Cen's progenitor: a nucleated dwarf}

In estimating the initial stellar mass of the nucleated dwarf 
(i.e., $\omega$ Cen's host galaxy),
we consider the following two points: (1) The mass fraction of the nucleus
(hereafter referred to as $f_{\rm n}$) 
is observed to range from 2 \% to 20 \% for nucleated dwarfs in nearby clusters
(e.g., Binggeli \& Cameron 1991) 
and (2) Given the observationally estimated  small
pericenter ($r_{\rm peri}$) and apocenter distance ($r_{\rm apo}$)
of the present-day  $\omega$ Cen's orbit with respect to the Galaxy  (Dinescu et al. 1999),
stripping of some fraction of stars 
through long-term ($\sim$ 10 Gyr) tidal interaction
with the Galaxy (e.g., Combes et al. 1999) is highly likely for $\omega$ Cen.
We therefore consider that the initial stellar mass 
$m_{\rm dw}$  
of its host is  equal to $(1.0-f_{\rm lost})^{-1}  \times f_{\rm n}^{-1} \times m_{\omega}$,
where $m_{\omega}$ is the present-day $\omega$ Cen's mass
(= 5.0 $\times$ $10^6$  M$_{\odot}$; Meylan  et al. 1995).  
If we adopt $f_{\rm lost}$ of 0.2  and $f_{\rm n}$ of 0.05, 
$m_{\rm dw}$ is 1.25$\times$ $10^8$ M$_{\odot}$.
The nucleated dwarf with  $M_{\rm B}$ ($B$-band absolute magnitude) 
is  assumed to consist  of dark matter halo, stellar envelope,
and nucleus. The stellar nucleus is modeled by the Plummer model with the scale
length of $a_{\rm n}$. 
Although we investigated both dwarf disc and spheroidal/elliptical models
for the stellar envelopes,
we show only the results of the dwarf disc models. 
We use the Fall-Efstathiou model (1980) for the dwarf disc models
with $M_{\rm t}$ = 10 $\times$ $M_{\rm d}$ 
and  the central $B-$band surface brightness (${\mu}_{0}$)
of 22 and  24 mag arcsec$^{-2}$.

\subsection{The dwarf's orbit}

The center of the Galaxy is set to be ($x$,$y$,$z$) = (0,0,0) and 
the initial position and  velocity of a dwarf are
($x$,$y$,$z$) = (cos$\theta$ $\times$ $r_{\rm in}$, 0, sin$\theta$ $\times$ $r_{\rm in}$)
and ($v_{\rm x}$,$v_{\rm y}$,$v_{\rm z}$) =  
(0, $v_{\rm in}$, 0), respectively, where $r_{\rm in}$, $\theta$, and, $v_{\rm in}$ 
are the distance from the Galactic center, the inclination angle of the dwarf's orbit
with respect to the Galactic plane, and the velocity of the dwarf, respectively. 
The positive sign of $v_{\rm in}$ represents a prograde  orbit
with respect to the Galactic rotation.
We present the results for the models  with $r_{\rm in}$ = 1.5 $R_{\rm d}$ (= 26.25 kpc),
in which the dwarf can intrude into the Galaxy from well outside the disc component.

\subsection{The fiducial  model}

We first searched  for a collisionless model in which the end product
satisfies the following two conditions:
(1) Only the dwarf's nucleus can survive with the envelope being almost completely stripped
and (2) the developed ``naked nucleus''  has  an orbit similar to that 
observed for the present-day $\omega$ Cen (i.e., strongly retrograde orbit
nearly confined within the Galactic  plane; e.g., Dinescu et al. 1999).
We found  that the  model with $M_{\rm B}$  $\sim$ $-14$ mag,
the dwarf's nucleus mass fraction of $\sim$ 0.05, 
${\mu}_{0}$ = 24 mag arcsec$^{-2}$,
$\theta$ $\sim$  30$^{\circ}$,  and $v_{in}$  $\sim$   $-$60 km s$^{-1}$  
(i.e., $e_{\rm p}$ $\sim$ 0.63)
can satisfy the above two required conditions in the bulgeless Galaxy model
with $M_{\rm d}$ = 0.2 $\times$ $M_{\rm d}(0)$  
and $F_{\rm d}$ = 10. 
We mainly show this fiducial 
results of the model.
Fig. 1 shows the orbital evolution for the fiducial model 
as well as for those which did not succeed in satisfying the two conditions.
These unsuccessful collisionless models suggest that   
(1) it is the density of the satellite that determines at which
galactiocentric radius it will disrupt, while it is the mass of
the satellite that determines the rate of orbital decay
and (2) the more massive young Galaxy and the Galaxy with the bulge 
can prevent the survival   of $\omega$ Cen-like clusters
with the pericenter of $<$ 5 kpc.

For the fiducial model, we also investigate the star formation history of the dwarf disc
by assuming the gas mass fraction of 0.1 and by adopting the Schmidt law (Schmidt 1959)
with an exponent of 1.5.
An  isothermal equation of state is used for the gas 
with a temperature of $2.5\times 10^3$ K.
We first describe the results of the collisionless fiducial model
and then describe the model including star formation.
The total number of particles used for each model
are 30000 for the Galactic dark matter, 20000 for the Galactic stellar disc,
10000 for the dwarf's dark matter,
20000 for the dwarf's stellar envelope, 10000 for the dwarf's gas,
and 5000 for the dwarf's nucleus (All of these are ``live''). 
All  the calculations related to the evolution of collisionless models
have been carried out on the GRAPE board (Sugimoto et al. 1990)
and the models including  star formation and hydrodynamical evolution 
are investigated by using  TREESPH codes described in  Bekki (1995, 1997).
Different gravitational softening lengths are allocated for different  components (e.g., nucleus)
so that we can investigate
the dynamical scale on  both the Galaxy-scale and the dwarf's scale.

\begin{figure}
\psfig{file=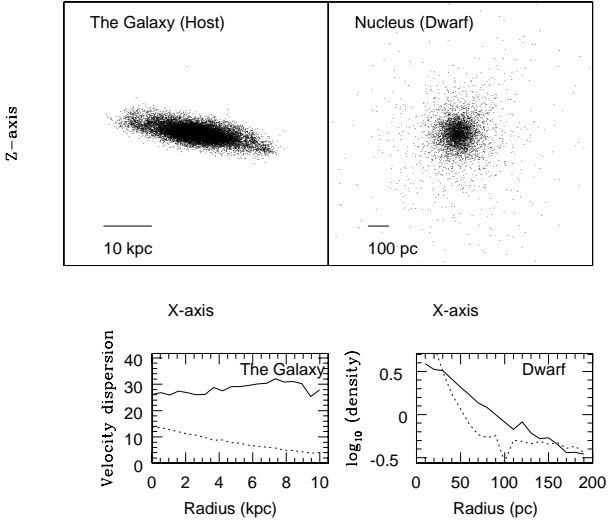,width=8.cm}
\caption{ 
{\it Upper two}: Final mass distribution of the Galactic stellar disc (left) 
and the survived dwarf's nucleus  (right) projected onto the $x$-$z$ plane 
at $T$ = 2.56 Gyr.
Note that the $\omega$ Cen's host dwarf can
cause significant dynamical heating (vertical thickening)
of the first generation of the Galactic 
thin disc.
{\it Lower two}: Initial (dotted) and final (solid, at $T$ = 2.56 Gyr) 
radial dependences of the vertical velocity
dispersion (${\sigma}_{\rm z}$) for  the initial Galactic disc in
units of km s$^{-1}$ (left) and the initial (dotted) and final (solid)
density distributions  of the dwarf  projected onto the $x-y$ plane
for the central 200 pc. The decrease of the central density of
the dwarf is due mostly to numerical effects caused by the introduction
of gravitational softening length. 
}
\label{Figure. 3}
\end{figure}

\section{Results}

Figs. 1 and 2 summarise the dynamical evolution of the stellar component
of the dwarf in the collisionless fiducial model.
As the dwarf sinks into the inner region of the Galactic disc owing to dynamical friction,
the outer low surface brightness stellar envelope is efficiently stripped by the Galactic strong
tidal force ($T$ = 0.93 Gyr).  
The compact prolate bar-like structure can be  formed  when the dwarf passes by
its orbital pericenter because of the  tidal perturbation.
The dwarf finally lose 
most of its initial stellar (and dark matter) mass within $\sim$ 2.6 Gyr.
The stripped stars forms an
inner stellar halo with the total mass of $\sim$ 10$^8$ $\rm M_{\odot}$ 
(corresponding to less than 10 \% of the present-day Galactic stellar halo mass)
and the initial Galactic thin disc finally become significantly thickened  
owing to the vertical  heating by the dwarf (See Fig. 3).

The nucleus, on the other hand, 
can survive  tidal destruction by the Galaxy
owing to its initial compact configuration (See Fig. 3).
The developed ``naked nucleus'' still follows the decayed dwarf orbit
at the time of dwarf's destruction so that it has orbital eccentricity of 0.78 
($r_{\rm peri}$ = 1.0 kpc and $r_{\rm apo}$ =8 kpc for the final 0.2 Gyr) 
similar to the observationally suggested  one ($\sim$ 0.7). 
The total mass within the central 100 pc of the dwarf
is 6.2 $\times$ $10^{6}$ $\rm M_{\odot}$ for the nucleus,
3.4 $\times$ $10^{6}$ $\rm M_{\odot}$ for the stellar envelope,
and 0.4 $\times$ $10^{6}$ $\rm M_{\odot}$ for the dark matter at $T$ = 2.6 Gyr,
which means that the surviving  nucleus is dominated by baryonic components.
Thus a  $\omega$ Cen-like stellar cluster with the mass of  
$10^{7}$ $\rm M_{\odot}$ and almost no dark matter  
can be formed from the dwarf galaxy dominated by dark matter.
We expect that this cluster  will  further slowly decrease its apocenter distance  
and its stellar mass 
owing to dynamical friction and tidal stripping (e.g., Zhao 2002). 

\section{Star formation history}

Fig. 4 shows that 
the star formation of the fiducial  model with gas dynamics 
is moderately enhanced around  $T$ $\sim$  1.2 Gyr 
(0.025 $\rm M_{\odot}$ yr$^{-1}$)
and $\sim$ 2.2 Gyr 
(0.02 $\rm M_{\odot}$ yr$^{-1}$; also  at $T$ = 0.2, 0.6, 0.8, 1.0, and 1.5 Gyr).
This enhancement of star formation can result from
the radial gas inflow induced by tidal torque
of the developed prolate (bar-like) stellar structure in the dwarf.
As a result of this, 
40 \% of the initial gas is consumed up by the triggered star formation,
and  mass fraction of new stars to the nucleus (old stars) 
within 200 pc becomes rather  high (0.21) at $T$ = 2.6 Gyr.

Since such efficient  star formation can naturally cause rapid chemical 
enrichment due to metal ejection from supernovae,
the new stars  are highly likely to be more metal-rich than
the old nucleus component and form a secondary peak of the metallicity distribution
of the nucleus.
These results suggest that the observed unique chemical evolution history 
of $\omega$ Cen
can be understand in the context of the repetitive radial gas inflow to 
the dwarf's nucleus
(and the resultant moderate starbursts and chemical enrichment there)
triggered by tidal interaction with the Galaxy. 
We also suggest that  the  observed spatial distribution and kinematics of relatively
metal-rich populations with $-1.2$ $\le$ [Fe/H] $\le$ $-0.6$ in $\omega$ Cen  
(Norris et al. 1997; Pancino et al. 2000; Ferraro et al. 2001)
can reflect the detail of the radial gaseous inflow processes in the dwarf nucleus.
The new stellar populations in the nucleus might well be  classified as 
new clusters rather than field stars,
because star clusters are more likely to be formed
in the central regions of interacting/merging galaxies
because of very high gas pressure (Bekki et al. 2002).
The merging of these young clusters with the dwarf's nucleus
can cause flattening (by rotation) of the shape of the nucleus  and
thus be responsible for the observed flattened shape 
and unusual rapid rotation of $\omega$ Cen.

\begin{figure}
\psfig{file=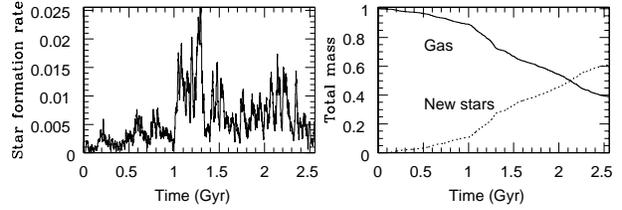,width=8.cm}
\caption{ 
The time evolution of star formation rate
(in units of $\rm M_{\odot}$ yr$^{-1}$)
in the fiducial model (left) and the time evolution
of the mass (normalized by the initial gas mass)
for the  gas  (solid) and the new stars (dotted)
in the model (right). 
}
\label{Figure. 4}
\end{figure}

\section{Discussions and conclusions}

If $\omega$ Cen was previously the nucleus of a  nucleated dwarf galaxy,
what fossil evidences for this can be seen in the Galactic halo region ?
Fig. 5 demonstrates that the tidally stripped stellar envelope
of the $\omega$ Cen's  host show a characteristic distribution 
in  the  $L_{\rm z} - L_{\rm xy}$ plane,
where $L_{\rm z}$ ($L_{\rm x}$, $L_{\rm y}$) and $L_{\rm xy}$ are
the angular momentum component in the $z$ ($x$, $y$) direction
and ${(L_{\rm x}^2+L_{\rm y}^2)}^{1/2}$, respectively.
These stars 
around the solar neighborhood also
show some crowding around $L_{\rm z}$ $\sim$ $-500$ and $L_{\rm xy}$ $\sim$ 300
kpc km s$^{-1}$, which reflects the orbital evolution of the dwarf.   
If we adopt the observed luminosity-metallicity relation 
${\rm [Fe/H]}_{\ast}=-3.43(\pm 0.14) - 0.157(\pm 0.012) M_{\rm V}$ 
(C\^ot\'e et al. 2000) for dwarfs,
we can expect that the stellar halo formed from the $\omega$ Cen's host
with $M_{\rm B}$ $\sim$ $-14$ mag
has the likely peak value of [Fe/H] $\sim$ $-1.2$  
(or somewhere between $-1.5$  and   $-0.84$ in [Fe/H])
in its metallicity distribution for $B-V$ = $0.5$.
Thus we suggest that  
the Galactic halo stars with [Fe/H] $\sim$ $-1.2$ and 
$L_{\rm z}$ $\sim$ $-500$ and $L_{\rm xy}$ $\sim$ 300 kpc km s$^{-1}$ 
can originate from $\omega$ Cen's  host. 


$\omega$ Cen-like objects have been already discovered in other galaxies and environments:
G1 in M31 (e.g., Meylan et al. 2001) and
very bright G1-like cluster in NGC 1023 (Larsen 2001).
We suggest that if $\omega$ Cen-like objects  in disc galaxies are formed from 
ancient nucleated dwarfs merging with  discs,  there should be some correlations between
the existence of $\omega$ Cen-like objects and structural properties of discs, 
because  galaxy interaction/merging can be responsible not only for the formation of 
thick discs and bars and for the formation of starbursts and AGNs (e.g., Noguchi 1987). 
For example, it is an interesting observational question whether or not disc galaxies with
$\omega$ Cen-like objects are more likely to have thick discs.

\begin{figure}
\psfig{file=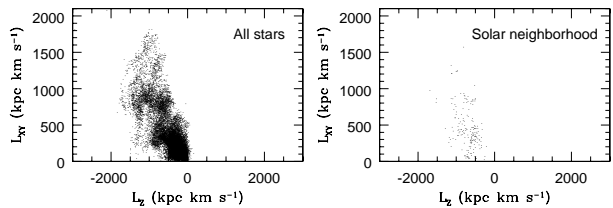,width=8.cm}
\caption{ 
The distribution of stars stripped from $\omega$ Cen's host dwarf
on the $L_{\rm z} - L_{\rm xy}$ plane for all stars (left)
and those within 5kpc from  the solar neighborhood (right).
Here  $L_{\rm z}$ ($L_{\rm x}$, $L_{\rm y}$) and $L_{\rm xy}$ are
the angular momentum component in the $z$ ($x$, $y$) direction
and ${(L_{\rm x}^2+L_{\rm y}^2)}^{1/2}$, respectively,
and this $L_{\rm xy}$ is not strictly a conserved quantity.
The right panel can be directly compared with observations 
shown in Fig. 15 of the paper  by
Chiba \& Beers (2000).
}
\label{Figure. 5}
\end{figure}

\section{Acknowledgment}
KB acknowledge the financial support of the Australian Research Council
throughout the course of this work.




\end{document}